\documentstyle[a4,proc,11pt]{article} 
\input amssym.def
\input amssym
\newcommand{\R}{\Bbb{R}}
\newcommand{\full}[1]{#1}

\newcommand{\conf}[1]{}

\newcommand{\squeezelist}{\setlength{\itemsep}{0pt}}

\def\e{{\epsilon}}
\def\P{{\cal P}}
\input psfig
\newenvironment{pf}{\unskip{\bf Proof:}}{\unskip{\hfill $\Box$}}

\newcommand{\lemlab}[1]{\label{lemma:#1}}

\newcommand{\tablab}[1]{\label{tab:#1}}
\newcommand{\figlab}[1]{\label{fig:#1}}
\newcommand{\seclab}[1]{\label{section:#1}}

\newcommand{\lemref}[1]{\ref{lemma:#1}}

\newcommand{\tabref}[1]{\ref{tab:#1}}
\newcommand{\figref}[1]{\ref{fig:#1}}
\newcommand{\eqref}[1]{(\ref{eq:#1})}
\newcommand{\secref}[1]{\ref{section:#1}}

\newtheorem{lemma}{Lemma}

{\catcode`\@=11
\gdef\setft#1#2#3{%
\def\@oddfoot{
{\setbox0=\hbox{#1}
\setbox1=\hbox{#3}
\ifdim\wd0>\wd1
\dimen0=\wd0
\box0\hfil#2\hfil\hbox to\dimen0{\hfil\hfil\box1}
\else \dimen0=\wd1
\hbox to\dimen0{\box0\hfil }\hfil#2\hfil\box1 \fi
}}} }

\def\complaint#1{}
\def\withcomplaints{
\newcounter{mycomplaints}
\def\complaint##1{\refstepcounter{mycomplaints}%
\ifhmode%
\unskip%
{\dimen1=\baselineskip \divide\dimen1 by 2 %
\raise\dimen1\llap{\tiny -\themycomplaints-}}\fi%
\marginpar{\tiny [\themycomplaints]: ##1}}%
}

\title{\bf Zero-Parity Stabbing Information
\conf{
\\
{\em (Abstract)}\\
\vspace*{-2mm}
}
}
\author{%
Joseph~O'Rourke and Irena~Pashchenko\thanks{
Dept. Comput. Sci., Smith Col\-lege, North\-ampton, 
MA 01063, USA.
\{orourke,\-ipashche\}@cs.\-smith\-.edu.
Supported by NSF grant CCR-9731804.}
}
\begin{document}
\maketitle
\vspace*{-18mm}
\begin{abstract}
\conf{\small}
\conf{
Everett et al.~\cite{ehn-sisp-96,full}\footnote{
	Bibliography omitted in this version.
}
} 
\full{Everett et al.~\cite{ehn-sisp-96,full}} 
introduced
several varieties of stabbing information for the lines
determined by pairs of vertices of a simple polygon $P$,
and established their relationships to vertex visibility
and other combinatorial data.
In the same spirit, we define the ``zero-parity (ZP) stabbing
information'' to be a natural weakening of their ``weak stabbing information,'' 
retaining only the distinction among
\{zero, odd, even $>0$\} in the number of polygon edges stabbed.
Whereas the weak stabbing information's relation to visibility
remains an open problem, we completely settle the analogous questions
for zero-parity information, with three results:
(1)~ZP information is insufficient to 
distinguish internal from external visibility
graph edges;
(2)~but it does suffice for all polygons that avoid a certain complex
substructure;
and (3)~the natural generalization of ZP information to the continuous
case of smooth curves does distinguish internal from external visibility.
\end{abstract}

\section{Introduction}
\vspace*{-2mm}
\conf{\small}
It is natural to connect the geometric shape of an object to its
combinatorics. 
The {\em polygon vertex visibility graph} has been
closely studied, but the relationship between this
graph and the shape remains open~\cite{o-cgc18-93}.
{\em Stabbing\/} information---how lines cross the polygon---has 
developed into a key concept both in discrete
geometry~\cite{g-gt-95,w-httgt-97} and 
geometric algorithmics~\cite{a-idapa-91,s-grp-97}.
The work of Everett et al.~\cite{ehn-sisp-96,full}
connects these two worlds, showing how
different varieties of stabbing information determine
visibility and other combinatorial information.

\conf{\small}
We now introduce enough notation to state our results.
Polygon vertices are assumed in general position 
and labeled by indices increasing
in a counterclockwise boundary traversal.  
The line $L$ through two vertices $x$ and $y$ of $P$ is partitioned
into three components $L \setminus \{x,y\}$,
and we count the number of edges of $P$ that cross each component:
Tail$(x,y)$,
Body$(x,y)$, and
Head$(x,y)$.
\full{
\vspace*{-6mm}
\begin{table}[htbp]
\conf{\small}
\begin{center}
\begin{tabular}{| c || c | c | c | c |}
        \hline
Stab Info
	& Cnv/Rfl 
	& Hull 
	& I/E Vis 
	& OrdTyp
	\\ \hline \hline
Labeled
	& {\sc yes}
	& {\sc yes}
	& {\sc yes}
	& {\sc yes}
	\\ \hline
Strong
	& {\sc yes}
	& {\sc yes}
	& {\sc yes}
	& {\sc no}
	\\ \hline
Weak
	& {\sc yes}
	& {\sc yes}
	& ?
	& {\sc no}
	\\ \hline \hline
Zero-Parity
	& {\sc yes}
	& {\sc yes}
	& {\sc no}
	& {\sc no}
	\\ \hline
Pure Parity
	& {\sc yes}
	& {\sc no}
	& {\sc no}
	& {\sc no}
	\\ \hline
\end{tabular}
\tablab{stab.results}
\end{center}
\conf{\small}
\end{table}
\vspace*{-5mm}
}
The {\em weak stabbing information\/} consists of these three quantities
for all pairs of vertices $(x,y)$.
Richer information leads to the {\em strong\/} and
{\em labeled\/} stabbing information,
which we will not pause to define.
We define {\em pure parity information\/} to only retain
the parity of Tail, Body, and Head.
As this is too weak to even identify hull 
\conf{edges,}
\full{edges (see Fig.~\figref{pure.parity}),
\begin{figure}[htbp]
\begin{center}
\ \psfig{figure=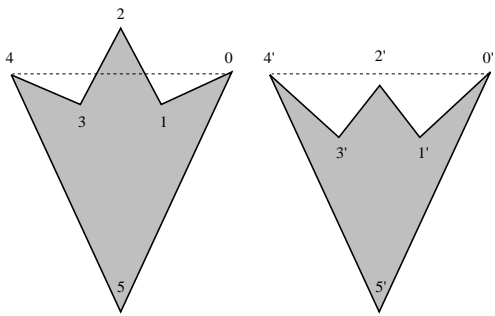,width=6.5cm}
\end{center}
\caption{Same pure parity information but different hull and visibility
edges.
}
\figlab{pure.parity}
\end{figure}
} 
we introduce
the {\em zero-parity (ZP) information\/}, which records three values:
zero (which enables visibility edges to be identified), odd,
and even $> 0$.
The results of~\cite{ehn-sisp-96,full} are summarized in
the first three lines of Table~\tabref{stab.results}, whereas
the last two lines display our contributions, completing the
table in a natural way.
\conf{
\vspace*{-6mm}
\begin{table}[htbp]
\small
\begin{center}
\begin{tabular}{| c || c | c | c | c |}
        \hline
Stab Info
	& Cnv/Rfl 
	& Hull 
	& I/E Vis 
	& OrdTyp
	\\ \hline \hline
Labeled
	& {\sc yes}
	& {\sc yes}
	& {\sc yes}
	& {\sc yes}
	\\ \hline
Strong
	& {\sc yes}
	& {\sc yes}
	& {\sc yes}
	& {\sc no}
	\\ \hline
Weak
	& {\sc yes}
	& {\sc yes}
	& ?
	& {\sc no}
	\\ \hline \hline
Zero-Parity
	& {\sc yes}
	& {\sc yes}
	& {\sc no}
	& {\sc no}
	\\ \hline
Pure Parity
	& {\sc yes}
	& {\sc no}
	& {\sc no}
	& {\sc no}
	\\ \hline
\end{tabular}
\tablab{stab.results}
\end{center}
\small
\end{table}
\vspace*{-10mm}
}

We concentrate on the ``I/E Vis'' column, distinguishing internal (I)
from external (E)
visibility edges.
It is natural to hypothesize that ZP information suffices to
make this distinction, due to the connection to the
well known ray-crossings point-in-polygon algorithm
\cite{h-pps-94}
\cite[Sec.~7.4]{o-cgc-98}, which depends only on parity.

\section{ZP Counterexample}
\vspace*{-2mm}
A counterexample to this hypothesis is shown in 
Fig.~\figref{zparity1}.\footnote{
	The three hull vertices $(9,10,11)$ are so far away that
	their 
	\conf{sight lines}\full{lines of sight}
	to the others are nearly vertical.
}
Let $[x,y]$ be the chain counterclockwise from $x$ to $y$.
The two $n=12$ vertex polygons differ in the subchains 
$[1,7]$
and $[1',7']$: the former lies below the 
I-edge $(0,8)$, and the latter above the E-edge $(0,8)$.
And yet all 
${12 \choose 2} \times 3 = 196$ pieces of
ZP information are identical (as checked by a program):
Tail$(1,6)=3$ and Tail$(1',6')=1$;
Head$(0,2)=3$ and Head$(0',2')=5$;
and so on.

\conf{\vspace*{-6mm}}
\begin{figure}[htbp]
\begin{center}
\ \psfig{figure=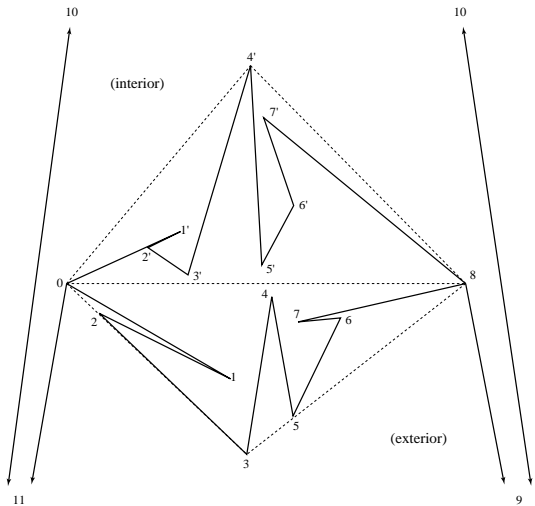,width=7cm}
\end{center}
\conf{\vspace*{-9mm}}
\caption{ZP counterexample.
}
\figlab{zparity1}
\end{figure}

\conf{\vspace*{-9mm}}
\section{Nontriangular Polygons}
\vspace*{-2mm}
The structures of the chains in Fig.~\figref{zparity1} are not
accidental, but rather the precise obstruction to distinguishing
I- from E-edges.
We prove that for any polygon that is {\em $({\ge}8)$-nontriangular},
ZP does suffice to make the distinction. 
Call a chain $[x,y]$ {\em $k$-triangular} if:
it contains $k$ vertices;
its hull is a triangle; and
$x$ and $y$ are seen by a vertex $z_I$ via I-edges, and by a
vertex $z_E$ via E-edges.
Note the $[0,8]$ chain in Fig.~\figref{zparity1} is $9$-triangular,
with $z_I=10$ and $z_E=9$ or $11$.
A polygon is $({\ge}k)$-nontriangular if it contains no $m$-triangular chain
for any $m \ge k$; 
thus it avoids all long triangular chains.
Our proof depends on a series of 
\conf{six lemmas, which we state informally.}
\full{six lemmas, whose proofs we only sketch.}
Let Even$(x,y,z)$ be the property that
$\forall w \in [x,y]$, Head$(w,z)$ is even,
and let Odd$(x,y,z)$ be the corresponding odd property.
\conf{
\begin{enumerate}
\squeezelist
\item The ZP information identifies convex/reflex and hull vertices.
\item Two vertices of a convex (resp.~reflex) chain are visible only
via an I-(resp. E)-edge.
\item If $xy$ is an I-edge, there is a 
convex $z$ in both chains bounded
by $x$ and $y$ such that Even$(x,y,z)$ holds
($10$ and $3/3'$ in Fig.~\figref{zparity1}.)
If $xy$ is an E-edge, there is a 
reflex $z$ in one of the chains
such that Odd$(x,y,z)$ holds
($4/4'$ in Fig.~\figref{zparity1}.)
\item If $xy$ is a nonhull I-(resp. E)-edge, it is shared by two
I-(resp. E)-$\triangle$s;
see Fig.~\figref{trilemma}.
\vspace*{-5mm}
\begin{figure}[htbp]
\begin{center}
\ \psfig{figure=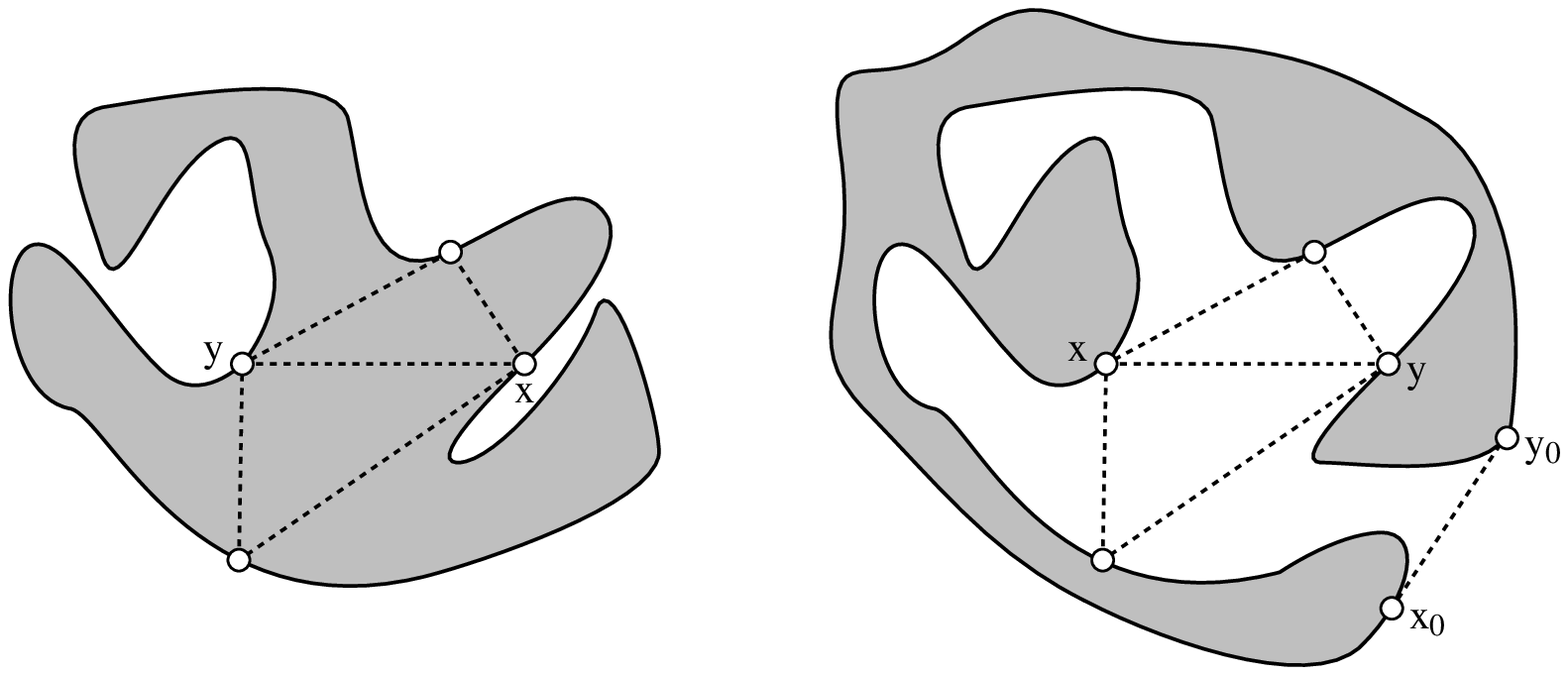,width=6.3cm}
\end{center}
\vspace*{-5mm}
\caption{$xy$ is an internal I- or E-edge.}
\figlab{trilemma}
\end{figure}
\item The previous lemmas I/E-dis\-tin\-guish all but visibility edges
spanned by triangular chains.
\item For a chain to be I/E-ambiguous, it must contain at least $8$ vertices.
(We believe this can be strengthened to $\ge 9$.)
\end{enumerate}
}
\full{
\begin{lemma}
The ZP information identifies convex/reflex and hull vertices.
\lemlab{one}
\end{lemma}
\begin{pf}
The proof of Lemma~4.1 in~\cite{full} carries through without change.
\end{pf}

\begin{lemma}
Two vertices of a convex (resp. reflex) chain are visible only
via an I-(resp. E)-edge.
\lemlab{two}
\end{lemma}
\begin{pf}
Suppose for a contradiction that
two vertices $x$ and $y$ of a convex chain $C$ are visible via
an E-edge.  Let $C' \subseteq C$ be the convex subchain partitioned
off by $xy$.  Orient $xy$ horizontal.  Then $C'$ must include a point
above or below $xy$.  Assume the latter.  
Then the rightmost
lowest point $v$ of $C'$ is a reflex vertex, a contraction.
\end{pf}

\begin{lemma}
If $xy$ is an I-edge, there is a 
convex $z$ in both chains bounded
by $x$ and $y$ such that Even$(x,y,z)$ holds
{\em ($10$ and $3/3'$ in Fig.~\figref{zparity1}.)}
If $xy$ is an E-edge, there is a 
reflex $z$ in one of the chains
such that Odd$(x,y,z)$ holds
{\em ($4/4'$ in Fig.~\figref{zparity1}.)}
\lemlab{three}
\end{lemma}
\begin{pf}
Orient the I-edge $xy$ horizontal,
and let $z_1$ be the highest vertex of $[x,y]$
and $z_2$ the lowest vertex of $[y,x]$.
Because the vertices $z_i$ are extreme, all rays through $z_i$
must exit the polygon there, so Even$(x,y,z_i)$ holds.
The reasoning for an E-edge is similar.
\end{pf}

\begin{lemma}
If $xy$ is a nonhull I-(resp. E)-edge, it is shared by two
I-(resp. E)-$\triangle$s.
\lemlab{four}
\end{lemma}
\begin{pf}
Let $xy$ be an I-edge. It must be part of a triangulation
of $P$.  By the nonhull assumption, $xy$
must be an internal diagonal and so shared by two triangles
of the triangulation;
see Fig.~\figref{trilemma}a.
The edges
of these triangles must be I-edges.
\begin{figure}[htbp]
\begin{center}
\ \psfig{figure=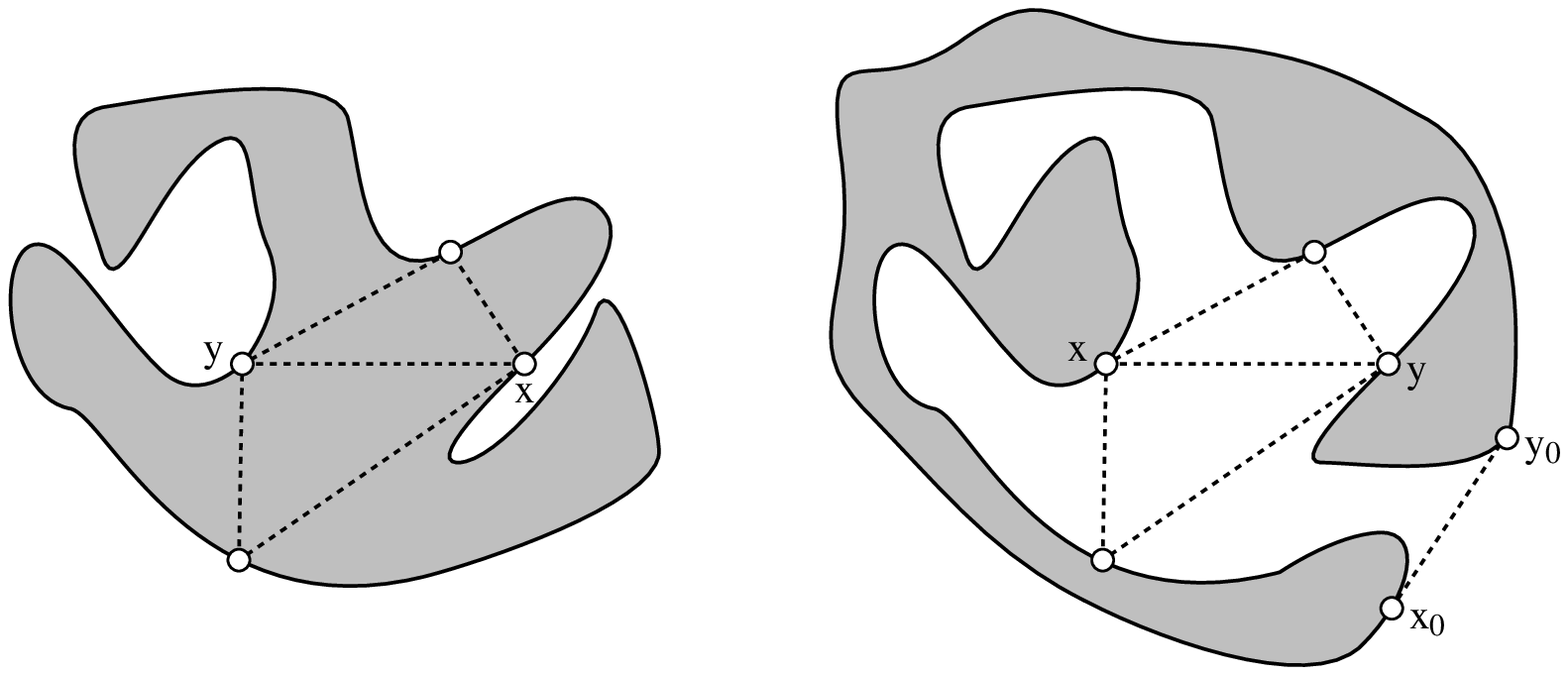,width=7cm}
\end{center}
\vspace*{-5mm}
\caption{$xy$ is an internal I- or E-edge.}
\figlab{trilemma}
\end{figure}
\end{pf}

\begin{lemma}
The previous lemmas
I/E-dis\-tin\-guish all but visibility edges
spanned by triangular chains.
\lemlab{five}
\end{lemma}
\begin{pf}
Suppose a nonhull visibility E-edge $xy$
satisfies
both the I- and E-halves of Lemma~\lemref{four}.
Then one of the two I-$\triangle$s that share $xy$ must
have its apex $z \in [x,y]$ (the other apex is in $[y,x]$).
With both $xz$ and $zy$ I-edges and $xy$ an E-edge, it
must be that the chain $[x,y]$ remains inside $\triangle xzy$.
Thus $z$ is on the hull, and $[x,y]$ is triangular.
\end{pf}

\begin{lemma}
For a chain to be I/E-ambiguous, it must contain at least $8$ vertices.
\lemlab{six}
\end{lemma}
\begin{pf}
Let $[x,y]$ be an I/E-ambiguous chain.  Lemma~\lemref{five}
shows it must be triangular regardless of whether $xy$ is
an I- or an E-edge.  The vertices on the hull must be
different in the two cases:  
a convex vertex $c$ if $xy$ is
an I-edge,
and a reflex vertex $r$ if an E-edge.
($3/3'$ and $4/4'$ respectively in Fig.~\figref{zparity1}.) 
Assume without loss of generality that the vertices occur
in the order $(x,c,r,y)$ in the I-chain, i.e.,
the one with $xy$ an I-edge; as in Fig.~\figref{zparity1},
we label the E-chain vertices with primes.
We now argue that there must be
at least two vertices between $x$ and $c$, and between $r$ and $y$.

From the point of view of $x$, 
the chain $[r',y]$ starts out below
the ray $xr'$ and ends up above the ray $xy$,
whereas 
the chain $[r,y]$ starts out below
the ray $xr$ and ends up again below the ray $xy$.
This difference demands at least one ``flip vertex'' $w$,
at which the polygon's relationship to the ray $xw$ differs
in the two chains. In Fig.~\figref{zparity1}, $w=6/6'$:
$5$ and $7$ are both below ray $06$ but $06'$ splits $5'$ and $7'$.
To achieve this difference with identical ZP information
requires in turn that the ray from $x$ lie on different sides
of the edge incident to $x$:
above $01$ but below $01'$ in the figure.
This forces this edge to aim so that it splits vertices
in $[r,y]$.
Ad hoc reasoning shows that neither of these can be $r$ or $y$,
so there must be two additional vertices in this chain.

Applying the same argument to $[x,c]$ from the point of view of $y$
leads to $6$ interior vertices, and so $8$ including $x$ and $y$.
\end{pf}

\noindent
We believe this lemma can be strengthened to $\ge 9$.
}

We have embodied these lemmas in a Java applet that accepts 
a user-specified polygon, 
computes the ZP information, and then classifies all visibility
edges as I or E, except for those spanned by triangular chains.
\full{For example, 
all $2566$ visibility edges in Fig.~\figref{irena.ex}
are correctly classified.
\begin{figure}[htbp]
\begin{center}
\ \psfig{figure=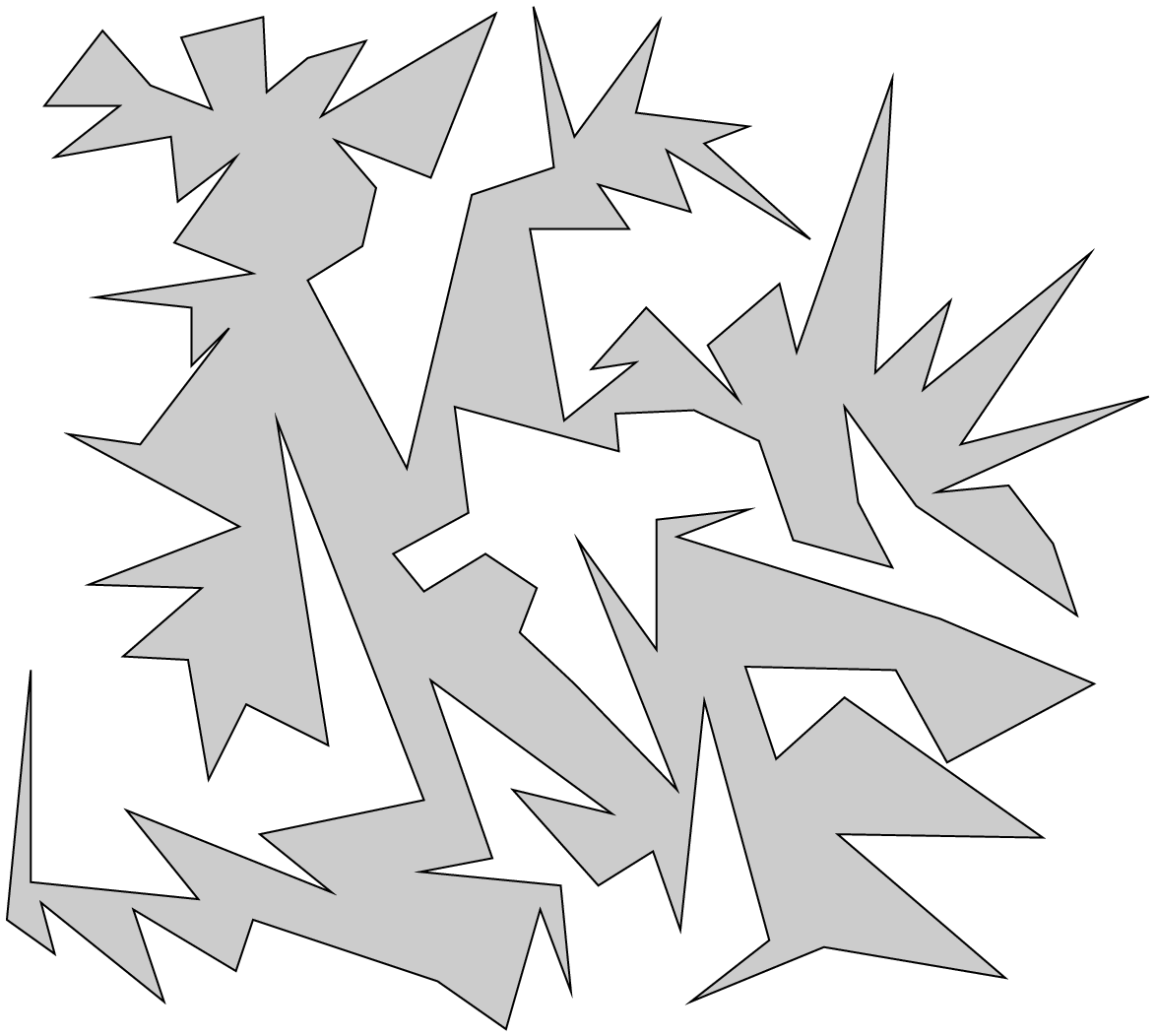,width=6cm}
\end{center}
\caption{A $({\ge}8)$-nontriangular polygon of $n=136$ vertices.}
\figlab{irena.ex}
\end{figure}
} 

\conf{
\vspace*{-3mm}
\section{Continuous ZP Info.}
} 
\full{
\vspace*{-2mm}
\section{Continuous ZP Info.}
} 
\seclab{continuous}
\vspace*{-2mm}
Shermer introduced the notion of {\em point visibility
graphs (PVGs)}~\cite{s-rrag-92,ms-isp-96},
a natural continuous generalization of vertex visibility graphs.
\conf{
We can generalize ZP information to continuous graphs and smooth
Jordan curves by associating with each point $x$, a function Body$_x(y)$
that records the cardinality of $xy \cap \partial P$;
and so on.\footnote{
	This definition requires that the curve 
	have nonzero curvature everywhere,
	so that every line meets the curve in a finite number of points.
}
The continuous ZP information can be finitely represented, as
it changes only at tangents as recorded in the 
visibility complex~\cite{pv-vc-96}.
We prove that the ZP functions permit classifying every PVG
edge as either an I- or an E-edge.
} 
\full{
We can generalize ZP information to continuous graphs and smooth
curves as follows.
Let $\P: [0,1] \rightarrow \R^2$ be a Jordan curve
parameterized by $t \in [0,1]$,
a piecewise algebraic curve
smooth except at no more than $n$ points,
and with nonzero curvature everywhere.
This latter condition ensures that 
every line meets the curve in a finite number of points.
The parameter $t$ plays the role of the vertex label.
For each point $x \in \P$, define a function
$B_x: [0,1] \rightarrow \{z,o,e\}$
so that $B_x(y)$ is the zero-parity of Body$(x,y)=|xy \cap \P|$.
Define $T_x()$ and $H_x()$ to similarly depend on
Tail$(x,y)$ and Head$(x,y)$.
The collection of these functions for all $x \in \P$ constitute
the continuous ZP information.

For a fixed $x$, each of the three functions $\{ B_x(),H_x(),T_x() \}$
is discontinuous only at points of tangency between the line through
$xy$ and $\P$.
The assumption that $\P$ is piecewise algebraic assures that
each function has at most $O(s n)$ discontinuities, where $s$ the maximum
degree of the algebraic pieces.  
And as $x$ varies over $\P$, 
the combinatorial structure of $B_x()$ changes only with double
tangencies, of which there are at most $O(s^2 n^2)$.
In fact, the visibility complex~\cite{pv-vc-96}
records all the relevant critical lines.
Thus the continuous ZP information
may be finitely represented.

Let $xy$ be a visibility edge, i.e., one for which $B_x(y) = z$.
The I/E status of $xy$ may be determined by examination of the ZP
functions in the local neighborhood of $xy$.
Care must be taken to deal with tangencies/discontinuities.
We label a discontinuity at $y$ by a value of the function
at $y-\e$, $y$, and $y+\e$, for small $\e > 0$: $o/z/e$, etc.

\begin{figure}[htbp]
\begin{center}
\ \psfig{figure=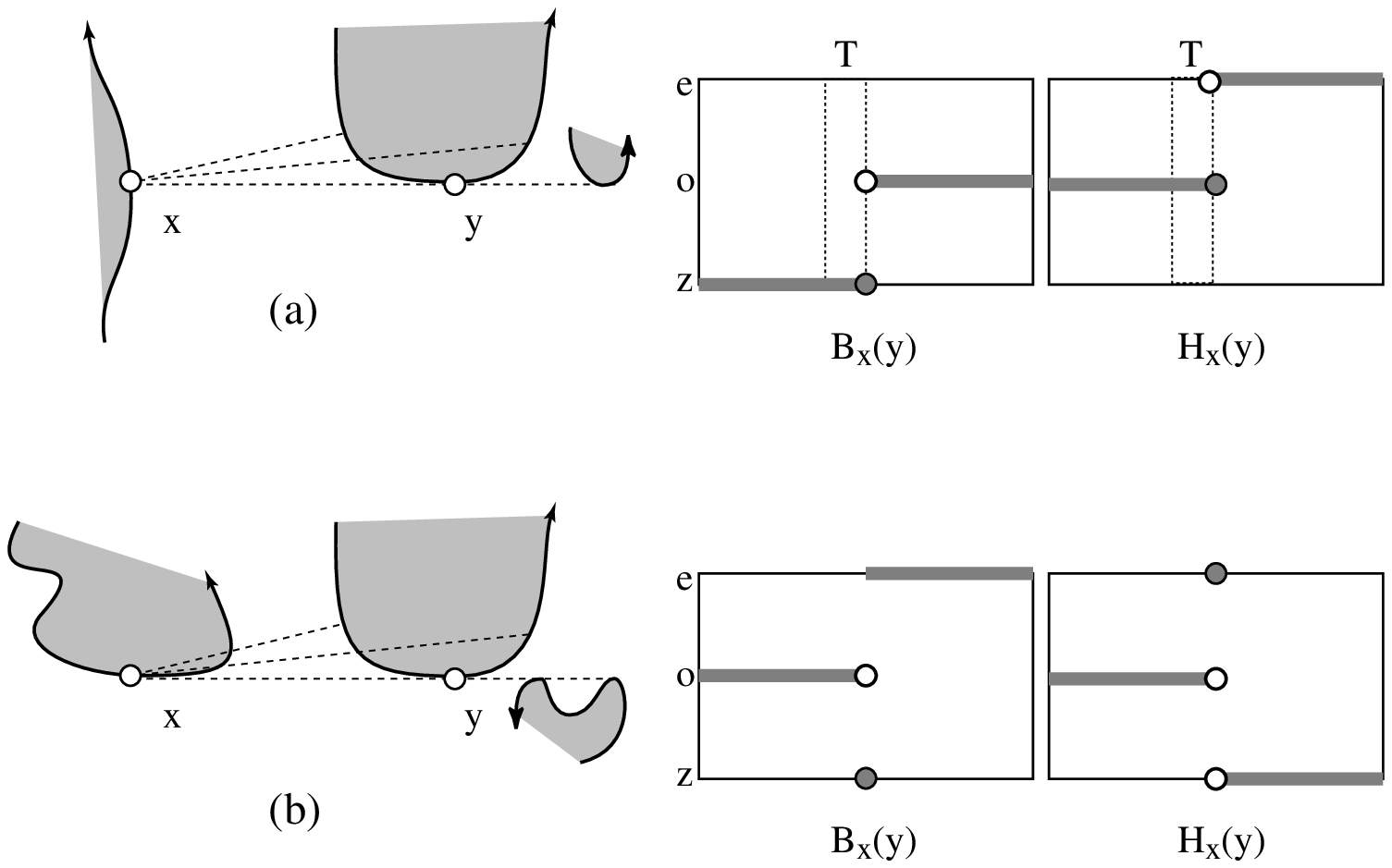,width=7.8cm}
\end{center}
\caption{Discontinuities in $B_x()$ and $H_x()$.}
\figlab{discont}
\end{figure}

\begin{lemma}
If $B_x(y)$ is either continuous at $y$, or has a
$z/z/o$ or $o/z/z$ discontinuity at $y$ 
{\em (see Fig.~\figref{discont}a)}, 
then the I/E status of $xy$ may be determined.
\lemlab{discont.0.1}
\end{lemma}
\begin{pf}
Let $T=(y,y + \delta)$ or $T = (y-\delta, y)$
be an interval incident to $y$ in which, for $t \in T$,
(a) both $B_x(t)$ and $H_x(t)$ are continuous,
and (b) $B_x(t) = z$.
Our assumptions guarantee such an interval.
Then,
\begin{enumerate}
\item[E:] If $H_x(t) = o$, $xy$ is an E-edge (Fig.~\figref{discont}a).
\item[I:] If $H_x(t) = z$ or $e$, $xy$ is an I-edge.
\end{enumerate}
When $B_x(y)$ is continuous at $y$, it can be shown
that $T$ on either side of $y$ leads to the same conclusion.
\end{pf}

\begin{lemma}
If $B_x(y)$ has a $o/z/e$
discontinuity at $y$, then $xy$ is an E-edge;
if it has an $e/z/o$ discontinuity, then $xy$ is an I-edge.
\lemlab{discont.2}
\end{lemma}
\begin{pf}
See Fig.~\figref{discont}b for an $o/z/e$
discontinuity.
Achieving $B_x(t) = e$ requires rays to intersect the curve $\P$ 
both near $x$ and near $y$.  The direction of the curve at $y$
is determined by the need to achieve $e$ after $y$.
This forces the direction of the curve at $x$ as shown; otherwise
we could not have $B_x(y)=z$.
The local situation then forces $xy$ to be an $E$ edge.
The $e/z/o$ discontinuity is the same with all directions
reversed.
\end{pf}

In addition it must be argued that the discontinuities covered by
the previous two lemmas are the only ones possible.
The final conclusion, that continuous ZP information distinguishes
I- from E-edges, justifies the intuition based on the point-in-polygon
algorithm.
} 

\conf{
\vspace*{-3mm}
\section{Weak Stabbing}
} 
\full{
\vspace*{-2mm}
\section{Open Problems}
} 
\vspace*{-2mm}
A judicious addition of two vertices to the chains in Fig.~\figref{zparity1}
produces a ``near'' counterexample to the hypothesis that the weak
stabbing information determines I/E, leaving ``only'' the Head$()$'s
and Tail$()$'s of $55$ vertex pairs to be equalized.
\full{
If this could be accomplished, the one `?' in Table~1 could be replaced
with {\sc no}.

The continuous stabbing information introduced in Section~\secref{continuous}
raises a number of new questions.
Generalization to Jordan curves with points of zero curvature (including
polygons) would be pleasing.
Connections to point X-ray theory~\cite[Ch.5]{g-gt-95} should be developed.
Identifying the equivalence class of curves that share the same
ZP information might be possible.
And it remains unclear what additional information is gained by
having the absolute stabbing numbers rather than only their zero-parity
information.
}

\conf{\newpage}
\small
\bibliographystyle{alpha}
\bibliography{stab,/home3/orourke/bib/geom/geom}
\end{document}